\documentclass[11pt]{article}
\usepackage{epsfig,cite}
\usepackage{graphicx}
\usepackage{multicol}
\usepackage{color}
\usepackage{amssymb}
\usepackage{amsthm}
\usepackage{amsmath,amsfonts}

\def\bea{\begin{eqnarray}}
\def\eea{\end{eqnarray}}

\def\be{\begin{equation}}
\def\ee{\end{equation}}

\newcommand\prd[3]   
		{{Phys.\ Rev.\ }{\bf D #1} (#2) #3}
\newcommand\prl[3]   
		{{Phys.\ Rev.\ Lett.\ }{\bf #1} (#2) #3}
\newcommand\plb[3]   
		{{Phys.\ Lett.\ }{\bf B #1} (#2) #3}
\newcommand\npb[3]    
		{{Nucl.\ Phys.\ }{\bf B #1} (#2) #3}
\newcommand\app[3]   
		{{Astropart.\ Phys.\ }{\bf #1} (#2) #3}
\newcommand\jhep[3]  
		{{J. High Energy Phys.\ }{\bf #1} (#2) #3}
\newcommand\epjc[3]  
		{{Eur.\ Phys.\ J. }{\bf C #1} (#2) #3}
\newcommand\npps[3]  
		{{Nucl.\ Phys.\ }{\bf #1} {\it(Proc.\ Suppl.)} (#2) #3}
\newcommand\jcap[3]  
		{{JCAP\ }{\bf #1} (#2) #3}

\newcommand\apj[3]    
		{{\it Astrophys.\ J. }{\bf #1} (#2) #3}

\newcommand{\lsim}{~{}_{\textstyle\sim}^{\textstyle <}~}

\def\sss{\scriptscriptstyle}

\begin{document}
\begin{titlepage}
\pagestyle{empty}
\baselineskip=21pt
\rightline{Caltech MAP-323}
\vspace*{2cm}
\begin{center}
{\huge\sf Supersymmetric Contributions to\\[0.4cm] Weak Decay Correlation Coefficients}\\[0.4cm] 
\end{center}
\begin{center}
\vskip 0.6in
{\Large\sf  S.~Profumo, M.~J.~Ramsey-Musolf and S.~Tulin}\\
\vskip 0.2in
{\it {California Institute of Technology, Pasadena, CA 91125, USA}}\\
{E-mail: {\tt profumo@caltech.edu, mjrm@caltech.edu, tulin@caltech.edu}}\\
\vskip 0.4in
{\bf Abstract}
\end{center}
\baselineskip=18pt \noindent

\noindent We study supersymmetric contributions to correlation coefficients that characterize the spectral shape and angular distribution for polarized $\mu$- and $\beta$-decays. In the minimal supersymmetric Standard Model (MSSM), one-loop box graphs containing superpartners can give rise to non-$(V-A)\otimes (V-A)$ four fermion operators in the presence of left-right or flavor mixing between sfermions. We analyze the present phenomenological constraints on such mixing and determine the range of allowed contributions to the weak decay correlation coefficients. We discuss the prospective implications for future $\mu$- and $\beta$-decay experiments, and argue that they may provide unique probes of left-right mixing in the first generation scalar fermion sector.

\vfill
\end{titlepage}
\baselineskip=18pt



\section{Introduction}

The search for physics beyond the Standard Model (SM) lies at the forefront of particle and nuclear physics. Among the prospective candidates for SM extensions, low-energy supersymmetry (SUSY) remains one of the most attractive possibilities. Its elegant solution to the naturalness problem associated with stability of the electroweak scale, its generation of coupling unification near the GUT scale, and its viable particle physics explanations for the abundance of matter (both visible and dark), have motivated a plethora of phenomenological studies over the years. With the advent of the Large Hadron Collider (LHC), direct evidence for low-energy SUSY may become available in the near future.

In the search for new physics, studies of precision electroweak observables and rare or SM-forbidden processes provide an important and complementary probe to collider searches (for recent discussions, see Refs.~ \cite{Erler:2004cx,Ramsey-Musolf:2006ur}). Indeed, precise measurements of $Z$-pole observables and other electroweak precision measurements, as well as the branching ratios of rare decays such as $b\rightarrow s\gamma$ or $B_s\rightarrow\mu^+\mu^-$ have placed important constraints on supersymmetric models. At low energies,  the recent evidence for a possibly significant deviation of the muon anomalous magnetic moment, $(g_\mu-2)$, from SM expectations provides at least a tantalizing hint of low-energy SUSY in the regime of large $\tan\beta$\cite{Bennett:2006fi,Czarnecki:2001pv}. Similarly, new searches for the permanent electric dipole moments of various systems will probe SUSY (and other) CP-violation sources at a level of interest to explaining the baryon asymmetry of the universe\cite{Pospelov:2005pr,Erler:2004cx}, while precision studies of fixed target, parity-violating electron scattering will be strongly sensitive to the existence lepton-number violating supersymmetric interactions\cite{Ramsey-Musolf:2005rz}. 

In this paper, we study the implications of SUSY for weak decays of the muon, neutron, and nuclei. Our work is motivated by the prospect of significantly higher precision in future measurements of the muon lifetime ($\tau_\mu$) and decay correlation parameters, as well as of considerably higher precision in studies of neutron and nuclear $\beta$-decay at various laboratories, including the Spallation Neutron Source (SNS), Los Alamos Neutron Science Center (LANSCE), NIST, the Institut  Laue-Langevin (ILL), and a possible high-intensity radioactive ion beam facility. Recent experimental progress including:  new measurements of muon decay parameters at TRIUMF and PSI\cite{Gaponenko:2004mi,Jamieson:2006cf, Danneberg:2005xv};  new measurements of the muon lifetime at PSI\cite{fast,mulan}; Penning trap studies of nuclear $\beta$-decay\cite{Hardy:2005qv}; the development of cold and ultracold neutron technology for the study of neutron decay; and plans for new measurements of the leptonic decays of the pion\cite{TRIUMFnew,PSInew} -- point to the high level of experimental activity in this direction. 

Theoretically, recent efforts have focused on the use of such experiments to test the unitarity of the Cabibbo-Kobayashi-Maskawa matrix, including analyses of SUSY corrections to the $(V-A)\otimes(V-A)$ structure of the SM charged current (CC) interaction 
\cite{Kurylov:2001zx,Ramsey-Musolf:2000qn,Barger:1989rk} and improved limits on hadronic structure effects in neutron and nuclear $\beta$-decays\cite{Marciano:2005ec}. Going beyond the $(V-A)\otimes (V-A)$ structure of the SM CC interaction, it has recently been shown that the scale of neutrino mass implied by neutrino oscillation experiments and the cosmic microwave background implies stringent bounds on chirality-changing scalar and tensor operators that could contribute to weak decays\cite{Ito:2004sh,Prezeau:2004md,Erwin:2006uc}. Comprehensive reviews of non-$(V-A)\otimes (V-A)$ effects in $\beta$-decay have been given in Refs.~\cite{Herczeg:2001vk,deutsch,Severijns:2006dr} 

Here, we study the effects of supersymmetric interactions that give rise to non-$(V-A)\otimes (V-A)$ interactions but evade the neutrino mass bounds. Such effects can arise through radiative corrections in supersymmetric models containing only left-handed neutrinos. 
For concreteness, we focus on the minimal supersymmetric Standard Model (MSSM). 
We do not consider simple extensions of the MSSM with right-handed or Majorana neutrinos and their superpartners as required by non-vanishing neutrino mass, as the  effects of the corresponding neutrino sector on weak decays is highly constrained.\footnote{In see-saw scenarios, for example, the scale of the additional sneutrino mass is sufficiently large that these degrees of freedom decouple from low-energy observables.}. We show that in the MSSM, radiatively-induced non-$(V-A)\otimes (V-A)$ interactions are particularly sensitive to flavor and left-right mixing among first and second generation sleptons and squarks and analyze the implications of such mixing for precision, low energy weak decay studies. In particular, we address the following question: if low-energy SUSY is the correct extension of the Standard Model and if the next generation of $\mu$- and $\beta$-decay studies uncover evidence for non-$(V-A)\otimes (V-A)$ interactions, what would be the corresponding implications for the MSSM?

The flavor structure of the MSSM has been the subject of extensive scrutiny, and as we show below, experimental studies of lepton flavor violation (LFV) lead to tight constraints on the corresponding effects in weak decays. In contrast, there exist few independent probes of left(L)-right(R) mixing among scalar fermions. This mixing, which occurs only after electroweak symmetry breaking (EWSB), receives two contributions proportional to one of the two neutral Higgs vacuum expectation values (vevs): (1) a supersymmetric term containing the product of the fermion Yukawa coupling, $Y_f$, and the bilinear Higgs superfield coupling, or $\mu$-parameter, and (2) a term proportional to the soft SUSY-breaking triscalar coupling, $a_f$. It is usually assumed that $|\mu | \lsim 1$ TeV in order to avoid fine-tuning in  obtaining proper EWSB. At the same time, adoption of the \lq\lq alignment " assumption for the soft triscalar coupling -- namely, that $a_f\propto Y_f$ -- provides a natural way to suppress flavor changing neutral currents, simplifies MSSM phenomenology, and -- as we discuss below -- constitutes a necessary ingredient for the additional Higgs scalars in the MSSM to be light enough to be accessible at future colliders. These two assumptions, which have guided SUSY phenomenology and model-building over the years, imply negligible L-R mixing among first and second generation sfermion superpartners. 

However, neither the assumption of electroweak scale $|\mu|$ or of soft triscalar alignment
is inherent in the MSSM superpotential or soft SUSY-breaking Lagrangian, and we may inhabit a world in which either one (or both) do not apply. Indeed, the degree of fine-tuning required for EWSB in the presence of large $|\mu|$ (up to the Planck scale) is far less severe than the fine tuning needed to obtain the observed small value of the cosmological constant, and arguments based on the avoidance of fine-tuning may not provide a fail safe guide to the scale of physical parameters. Similarly, we do not yet possess any direct experimental evidence for the even the SM Higgs boson, let alone the additional Higgs scalars implied by the MSSM, so requiring that the triscalar couplings be small enough to ensure that these scalars are light enough to be seen at the LHC or ILC may reflect more of a theoretical prejudice than a requirement of experimental observation. In short, assuming negligible L-R mixing for first and second generation sfermions is neither a first principles theoretical requirement or phenomenological necessity, and it is of interest to explore experimental tests of this assumption. In what follows, we argue that studies of weak decay correlations may provide such experimental tests. Specifically, we find that:

\begin{itemize} 

\item[(i)] Supersymmetric box graph corrections to the $\mu$-decay amplitude generates a non-vanishing scalar interaction involving right handed charged leptons ($g^S_{RR}\not=0$ in the standard parameterization used below) in the presence of flavor mixing among left-handed (LH) sneutrinos and among right-handed (RH) sleptons, or flavor-diagonal mixing among LH and RH sleptons.

\item[(ii)] Analogous box graph effects can give rise to non-vanishing scalar and tensor interactions in light quark $\beta$-decay. The generation of these interactions requires non-vanishing left-right mixing among first generation sleptons and squarks. Studies of the energy-dependence of $\beta$-decay correlations and $\beta$-polarization -- as well as of the energy-independent spin-polarization correlation --  provide a probe of these interactions and the requisite L-R mixing.

\item[(iii)] Flavor-mixing among the LH sleptons (${\tilde\nu}_L$, ${\tilde\ell}_L$) is highly constrained by searches for lepton flavor violation in processes such as $\mu\to e\gamma$ and $\mu\to e$ conversion. Thus, any observable departure from $(V-A)\otimes(V-A)$ interactions in $\mu$-decay associated with the MSSM would arise from large, flavor diagonal  L-R mixing among smuons ($\tilde\mu$)  and selectrons ($\tilde e$). The former are also constrained by the present value of $(g_\mu-2)$. At present, we are not aware of any analogous constraints on L-R mixing among first generation squarks and sleptons.

\item[(iv)] The magnitude of the effects in $\mu$-decay are below the present sensitivity of decay correlation studies. However, the presence of the SUSY-induced scalar interaction  could modify the extraction of the Fermi constant ($G_\mu$) from the next generation of $\tau_\mu$ measurements at PSI. Similarly, improvements in $\beta$-decay correlation precision by $\lesssim$ an order of magnitude  would allow one to probe the SUSY-induced scalar and tensor interactions generated by large L-R mixing. Such measurements could in principle provide a unique test of L-R mixing among first generation superpartners. 

\item[(v)] The observation of scalar and tensor interactions in low-energy weak decays would have severe implications for either the Higgs sector of the MSSM, the conventional description of EWSB with a weak scale $\mu$ parameter, or both. Although a detailed discussion of the Higgs sector of the MSSM goes beyond the scope of the present study -- where we focus on the low-energy phenomenology of supersymmetric radiative corrections -- we suggest possible directions for future theoretical exploration of a super heavy SUSY Higgs sector.

\end{itemize}

In the remainder of the paper, we provide details of our analysis. Section \ref{sec:general} gives a general overview of weak decay correlations and our notation and conventions. In Section \ref{sec:scalar} we discuss the computation of the relevant SUSY corrections and give analytic expressions for the resulting operators. Section \ref{sec:constraints} contains a discussion of constraints resulting from other measurements and numerical implications for the $\mu$-decay and $\beta$-decay correlations.  We summarize our conclusions in Section \ref{sec:conclude}.

\section{Weak Decay Correlations: General Features}
\label{sec:general}

Departures from the SM $(V-A)\otimes (V-A)$ structure of the low-energy leptonic and semileptonic CC weak interactions can be characterized by an effective four fermion Lagrangian containing all independent dimension six operators. In the case of $\mu$-decay, it is conventional  to use
\begin{equation}
\label{eq:leffmu}
{\cal L}^{\mu-\rm decay} = - \frac{4 G_\mu}{\sqrt{2}}\ \sum_{\gamma,\, \epsilon,\, \mu} \ g^\gamma_{\epsilon\mu}\, 
\ {\bar e}_\epsilon \Gamma^\gamma \nu_e\, {\bar\nu}_\mu \Gamma_\gamma \mu_\mu
\end{equation}
where the sum runs over Dirac matrices $\Gamma^\gamma= 1$ (S), $\gamma^\alpha$ (V), and $\sigma^{\alpha\beta}/\sqrt{2}$ (T) and the subscripts $\mu$ and $\epsilon$  denote the chirality ($R$,$L$) of the muon and final state lepton, respectively\footnote{The normalization of the tensor terms corresponds to the convention adopted in Ref.~\cite{Scheck}}. Note that the use of this Lagrangian is only appropriate for the analysis of processes that occur at energies below the electroweak scale, as it is not SU(2)$_L\times$U(1)$_Y$ invariant. At tree-level in the SM one has $g^V_{LL}=1$ with all other $g^\gamma_{\epsilon\mu}=0$. In the limit of vanishing lepton masses, non-QED SM electroweak radiative corrections to the tree-level amplitude are absorbed into the definition of $G_\mu$.

In the literature, there exist several equivalent parameterizations of non-Standard Model contributions to light quark $\beta$-decay\cite{Herczeg:2001vk,deutsch,Severijns:2006dr}. In analogy with Eq.~(\ref{eq:leffmu}) we use
\begin{equation}
\label{eq:leffbeta}
{\cal L}^{\beta-\rm decay} = - \frac{4 G_\mu}{\sqrt{2}}\ \sum_{\gamma,\, \epsilon,\, \delta} \ a^\gamma_{\epsilon\delta}\, 
\ {\bar e}_\epsilon \Gamma^\gamma \nu_e\, {\bar u} \Gamma_\gamma d_\delta
\end{equation}
where the notation is similar to that for the $\mu$-decay effective Lagrangian. As with ${\cal L}^{\mu-\rm decay}$, only the  purely left-handed $(V-A)\otimes (V-A)$ interaction appears at tree-level in the SM. In this case, one has $a^V_{LL}=V_{ud}$, the (1,1) element of the Cabibbo-Kobayashi-Maskawa (CKM) matrix. Including electroweak radiative corrections leads to
\be
a^V_{LL}=V_{ud}\, \left(1+{\Delta\widehat r}_\beta-{ \Delta\widehat r}_\mu\right)\ \ \ ,
\ee
where ${\Delta \widehat r}_\beta$ contains the electroweak radiative corrections to the tree-level $(V-A)\otimes (V-A)$ $\beta$-decay amplitude and ${\Delta \widehat r}_\mu$ contains the corresponding corrections for $\mu$-decay apart from the QED corrections that are explicitly factored out when definiting $G_\mu$ from the muon lifetime. 

Supersymmetric contributions to ${\Delta \widehat r}_\beta-{\Delta \widehat r}_\mu$ have been computed in Refs.~\cite{Kurylov:2001zx,Ramsey-Musolf:2000qn,Barger:1989rk}. These corrections can affect tests of the unitarity of the first row of the CKM matrix, as they must be subtracted from $a^V_{LL}$ when determining $V_{ud}$ from $\beta$-decay half lives. The corresponding implications for CKM unitarity tests have been discussed in those studies. Supersymmetric corrections can also give rise to non-vanishing $g^\gamma_{\epsilon\mu}$ and $a^\gamma_{\epsilon\delta}$ that parameterize the non-$(V-A)\otimes(V-A)$ interactions in Eqs.~(\ref{eq:leffmu},\ref{eq:leffbeta}). The presence of these operators cannot be discerned using the muon lifetime or $\beta$-decay half lives alone, but they can be probed using studies of the spectrum, angular distribution, and polarization of the decay products.  We consider here non-$(V-A)\otimes(V-A)$ that are generated by one-loop corrections and are, thus, suppressed by a factor of $\alpha/4\pi$. Consequently, we focus on those operators that interfere linearly with SM contributions in weak decay observables and have the largest possible phenomenological effects.

In the case of polarized $\mu^-$ ($\mu^+$) decay, the electron (positron) spectrum and polarization are characterized by the eleven Michel parameters\cite{michel1,michel2}, four of which ($\rho$, $\eta$, $\xi$, and $\delta$) describe the spectral shape and angular distribution. An additional five ($\xi^\prime$, $\xi^{\prime\prime}$, $\eta^{\prime\prime}$, $\alpha/A$, $\beta/A$) are used to characterize the electron (positron) transverse and longitudinal polarization, while the final two ($\alpha^\prime/A$, $\beta^\prime/A$) parameterize time-reversal odd correlations between the muon polarization and the outgoing charged lepton spin.  In what follows, we find that SUSY box graphs generate non-vanishing contributions to $g^S_{RR}$. This coupling appears quadratically in the parameters $\xi$ and $\xi^\prime$ and interferes linearly with the SM term $g^V_{LL}$ in $\eta$, $\eta^{\prime\prime}$, and $\beta^\prime/A$. The linear-dependence of $\eta$ on $g^S_{RR}$ is particularly interesting, since $\eta=0$ in the SM, and since a non-zero value for this parameter enters the extraction of $G_\mu$ from $\tau_\mu$:
\be
\label{eq:mudecayrate}
\frac{1}{\tau_\mu}=\frac{G_\mu^2 m_\mu^5}{192\pi^3} \left[1+\delta_{\rm QED}\right]\left[1+4\eta\frac{m_e}{m_\mu}-8\left(\frac{m_e}{m_\mu}\right)^2\right]\left[1+\frac{3}{5}\left(
\frac{m_\mu}{M_W}\right)^2\right]\ \ \ ,
\ee
where $\delta_{\rm QED}$ denote the QED corrections the decay rate in the low-energy (Fermi) effective theory. 

For $\beta$-decay, it is customary to use the description of the differential decay rate written down by Jackson, Treiman, and Wyld \cite{jtw} (see also \cite{Herczeg:2001vk,deutsch,Severijns:2006dr}):
\begin{eqnarray}
\label{eq:betacor}
   d\Gamma& \propto & {\cal N}(E_e)\Biggl\{ 1+a {{\vec p}_e\cdot{\vec p}_\nu\over E_e E_\nu}
   + b{\Gamma m_e\over E_e} + \langle {\vec J}\rangle\cdot \left[A{{\vec p}_e\over E_e} 
   + B{{\vec p}_\nu \over E_\nu} + D{{\vec p}_e\times {\vec p}_\nu \over E_e E_\nu}\right] \\
 \nonumber
&&+ {\vec\sigma}\cdot\left[N \langle{\vec J}\rangle + G\frac{{\vec p}_e}{E_e}+Q^\prime {\hat p}_e {\hat p}_e\cdot \langle{\vec J}\rangle+R \langle {\vec J}\rangle\times\frac{{\vec p}_e}{E_e}\right]
 \Biggr\}
   d\Omega_e d\Omega_\nu d E_e,
\end{eqnarray}
where ${\cal N}(E_e)=p_e E_e(E_0-E_e)^2$; $E_e$ ($E_\nu$), ${\vec p}_e$ 
(${\vec p}_\nu$), and ${\vec\sigma}$ are the $\beta$ (neutrino) energy, momentum, and polarization, respectively; ${\vec J}$ is the  polarization of the decaying nucleus; and $\Gamma=\sqrt{1-(Z\alpha)^2}$. 

As we discuss below, SUSY box graphs may generate non-zero contributions to the operators in Eq.~(\ref{eq:leffbeta}) parameterized by $a^S_{RR}$, $a^S_{RL}$, and $a^T_{RL}$. The latter interfere linearly with the SM parameter $a^V_{LL}$ in 
terms in Eq.~(\ref{eq:betacor}) that depend on $\beta$ energy, including the so-called Fierz interference coefficient, $b$; the parity-violating correlation involving neutrino momentum and nuclear spin, $B$; and the polarization correlation coefficient, $Q^\prime$.  In addition, the energy-independent spin-polarization correlation coefficient $N$ also contains a linear dependence on $a^S_{RR}$, $a^S_{RL}$, and $a^T_{RL}$. Specifically, one has 
\begin{align}
b\:\zeta =& \: \pm \; 4 \: {\textrm{Re}}\left[ M_F^2 \: g_V g_S \: a^V_{LL} (a^S_{RL}+a^S_{RR})^* -
                            2 M_{GT}^2 \: g_A g_T \: a^V_{LL} a^{T*}_{RL} \frac{}{} \right] \\
B\:\zeta =& \; 2 \; {\textrm{Re}} \left[ \: \pm \: \lambda_{J^\prime J} \: M_{GT}^2 \: 
              \left( (g_A^2 \: |a^V_{LL}|^2 + 4 \: g_T^2 \: |a^T_{RL}|^2) \; 
                      \mp \; \frac{\Gamma m}{E} \; 4 \: g_A g_T \: a^V_{LL} a^{T*}_{RL} \right) \right. \\
& \; \quad \qquad 
  + \delta_{J^\prime J} \; M_F M_{GT} \: \sqrt{\frac{J}{J+1}} \; \left( 2 \: g_V g_A \: |a^V_{LL}|^2 \frac{}{} \right. 
\notag \\
& \qquad \qquad \qquad \qquad \mp \; \left. \frac{\Gamma m}{E} \: 
(4 \: g_V g_T \: a^V_{LL} a^{T*}_{RL} - 
2 \: g_A g_S \: a^V_{LL}(a^S_{RL}+a^S_{RR})^* ) \left. \frac{}{} \right) \right] 
\notag \\
Q^\prime \; \zeta =& \; 2 \; {\textrm{Re}} \left[ \: \lambda_{J^\prime J} \: M_{GT}^2 \: 
              \left( (g_A^2 \: |a^V_{LL}|^2 + 4 g_T^2 \: |a^T_{RL}|^2) \; 
                      \mp \; \frac{\Gamma m}{E} \; 2 \:g_A g_T \: a^V_{LL} a^{T*}_{RL} \right) \right. \\
& \; \quad \qquad 
\mp \; \delta_{J^\prime J} \; M_F M_{GT} \: \sqrt{\frac{J}{J+1}} \; \left( 2 \: g_V g_A \: |a^V_{LL}|^2 \frac{}{} \right. 
\notag \\
& \qquad \qquad \qquad \qquad \mp \; \left. \frac{\Gamma m}{E} \; (4 \: g_V g_T \: a^V_{LL} a^{*T}_{RL}
                - 2 \: g_A g_S \: a^V_{LL} (a^S_{RL}+a^S_{RR})^*  ) \left. \frac{}{} \right) \right] \notag\\
N \; \zeta =& \; 2 \: {\textrm{Re}} \left[ \: \lambda_{J^\prime J}  \; M^2_{GT} \; \left( \frac{\Gamma m}{E} \;
  (4 g_T^2 \: |a_{RL}^T|^2 + g_A^2 \: |a^V_{LL}|^2 ) \mp 4 g_T g_A \: a^T_{RL} a^{V*}_{LL} \right) \right. \\
& \; \quad \qquad
+ \; \delta_{J^\prime J} \; M_F M_{GT} \; \sqrt{\frac{J}{J+1}} \: 
    \left( (4 g_V g_T \: a^V_{LL} a^{T*}_{RL} - 2 g_S g_A \: (a^S_{RR}+a^S_{RL})a^{V*}_{LL}) \frac{}{} \right. \notag \\
& \qquad \qquad \qquad \qquad \pm \;  \left. \frac{\Gamma m}{E} \; 
  (4 g_S g_T \: (a^S_{RR} + a^S_{RL}) a^{T*}_{RL} - 2 g_V g_A \: |a^V_{LL}|^2 ) \left. \frac{}{} \right) \right] \notag \\
\zeta =& \; 2 M_F^2 \: \left( g_V^2 \: |a^V_{LL}|^2 + g_S^2 \: |a^S_{RL}+a^S_{RR}|^2 \right) + 
      2 M_{GT}^2 \: (g_A^2 \: |a^V_{LL}|^2 + 4 \: g_T^2 \: |a^T_{RL}|^2)\ \ \ ,\\
\end{align}
where $J$ ($J'$) are the initial (final) nuclear spin and $\lambda_{J^\prime J}=(1, 1/J+1, -J/J+1)$ for
$J^\prime = (J-1, J, J+1)$\footnote{The quantity $\zeta$ is often denoted by $\xi$ in the literature. However, we have modified the notation to avoid confusion with the Michel parameter $\xi$.}. 
The quantities
$M_F$ and $M_{GT}$ are Fermi and Gamow-Teller matrix elements and $g_{V,A,S,T}$ are vector, axial vector, scalar, and tensor form factors. For transitions between initial ($i$) and final ($f$) nuclear states the corresponding reduced matrix elements are
\begin{eqnarray}
\nonumber
\langle f || {\bar u}\gamma^\lambda d + {\rm H.C.} || i \rangle & = & g_V(q^2) M_F \\
\langle f || {\bar u}\gamma^\lambda \gamma_5 d + {\rm H.C.} || i \rangle & = & g_A(q^2) M_{GT} \\
\nonumber
\langle f || {\bar u} d + {\rm H.C.} || i \rangle & = & g_S(q^2) M_F \\
\nonumber
\langle f || {\bar u}\sigma^{\lambda\rho} d + {\rm H.C.} || i \rangle & = & g_T(q^2) M_{GT}\ \ \ .
\end{eqnarray}
The conserved vector current property of the SM CC interaction implies that $g_V(0)=1$ in the limit of exact isospin symmetry. Isospin breaking corrections imply deviations from unity of order a few $\times 10^{-4}$ \cite{Kaiser:2001yc} (for an earlier estimate, see Ref.~\cite{Behrends:1960nf}). A two parameter fit to $\beta$-decay data yields $g_A(0)/g_V(0) = 1.27293(46)$\cite{Severijns:2006dr}, assuming only a non-vanishing SM coupling, $a^V_{LL}$, and neglecting differences in electroweak radiative corrections between hadronic vector and axial vector amplitudes. Theoretical expectations for $g_S$ and $g_T$ are summarized below.

\section{SUSY-Induced Scalar and Tensor Interactions}
\label{sec:scalar}

\begin{figure}[!t]
\begin{center}
\epsfig{file=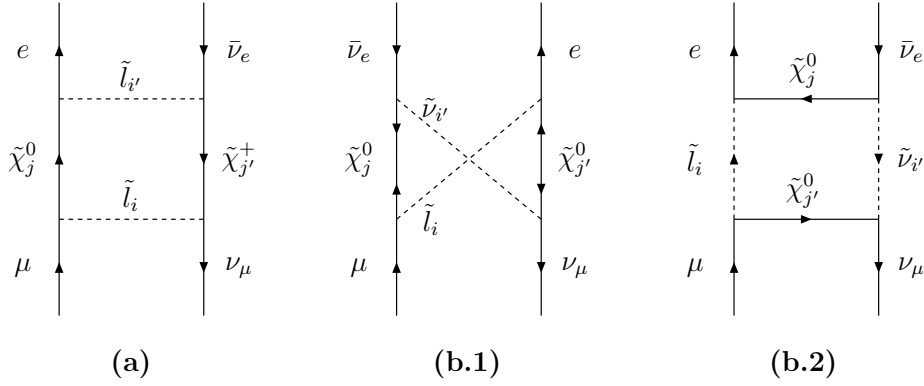,height=5cm}
\end{center}
\caption{\it\small 
Feynman diagrams relative to supersymmetric contributions giving rise to non $(V-A)\otimes(V-A)$ amplitudes in the muon decay. The amplitude relative to the diagram shown in {\bf (a)} involves left-right slepton mixing, while those in {\bf (a)} are non-vanishing if lepton flavor mixing is present in the slepton sector.}
\label{fig:feyn_mu}
\end{figure}
\begin{figure}[!h]
\begin{center}
\epsfig{file=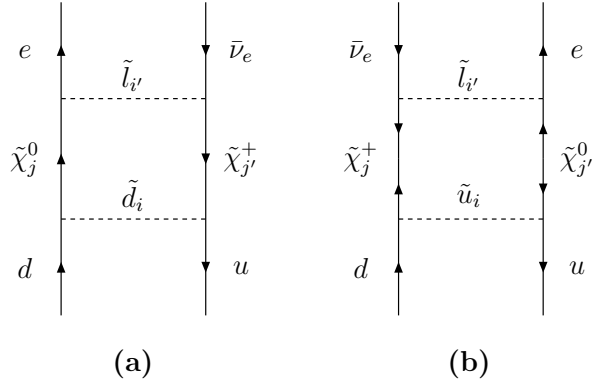,height=5cm}
\end{center}
\caption{\it\small 
Feynman diagrams relative to supersymmetric contributions giving rise to anomalous amplitudes in $\beta$ decay processes.}
\label{fig:feyn_beta}
\end{figure}
We compute the SUSY contributions to the weak decay correlations in the MSSM and obtain non-vanishing contributions to $g^S_{RR}$, $a^S_{RR}$,  and $a^{S,T}_{RL}$ from the box diagrams in Figures \ref{fig:feyn_mu} ($\mu$-decay) and \ref{fig:feyn_beta} ($\beta$-decay). These amplitudes -- as well as others not shown explicitly in Figures \ref{fig:feyn_mu} and \ref{fig:feyn_beta} -- also contribute to the parameters $g^V_{LL}$ and $a^V_{LL}$ that arise in the SM. A complete analysis of those contributions, along with the gauge boson propagator, vertex, and external leg corrections, is given in Ref.~\cite{Kurylov:2001zx}. Here, we focus on the non-$(V-A)\otimes(V-A)$ operators generated by the diagrams shown explicitly. 

Since the SM CC interaction is purely left-handed (LH), the generation of operators involving the right-handed (RH) SM fields requires the presence of RH fermion superpartners in the one-loop graphs. These particles appear either by virtue of left-right mixing among superpartners in Figures 
\ref{fig:feyn_mu}(a) and \ref{fig:feyn_beta}(a,b) or through coupling of the neutralinos ($\tilde\chi^0_j$) to the RH sleptons as in Figures \ref{fig:feyn_mu}(b.1, b.2). Note that in the latter case, a given virtual slepton mass eigenstate $\tilde l_i$ will couple to both the first and second generation charged leptons, thereby requiring the presence of non-zero flavor mixing. In contrast, the contributions of Figures. \ref{fig:feyn_mu}(a) and \ref{fig:feyn_beta}(a,b) involve only L-R mixing but no flavor mixing among the sfermions in the loops.

To set our notation, we largely follow the conventions of Refs.~\cite{Rosiek:1989rs,Rosiek:1995kg}. The L-R and flavor mixing among the sfermions is determined by the sfermion mass matrices. In the flavor basis, one has 
\be
{\bf M_{\tilde f}^2} =\left(
\begin{array}{cc}
{\bf M_{LL}^2} & {\bf M_{LR}^2}\\
{\bf M_{LR}^2} & {\bf M_{RR}^2}
\end{array}\right)
\ee
where for quark and charged slepton superpartners ${\bf M_{AB}^2}$  ($A,B=L,R$) are $3\times 3$ matrices with indices running over the three flavors of sfermion of a given chiral multiplet ($L,R$). For sneutrinos, only $ {\bf M_{LL}^2}$ is non-vanishing. After electroweak symmetry-breaking, the 
${\bf M_{AB}^2}$ take the forms (using squarks as an illustration)
\begin{equation}
\label{eq:mll}
{\bf{M_{LL}^2}}  =    {\bf m_Q^2}+ {\bf m_q^2 }+{\bf \Delta_f} 
\end{equation}
\begin{equation}
\label{eq:mrr}
{\bf{M_{RR}^2}}  =    {\bf m_{\bar f}^2}+ {\bf m_q^2 }+{\bf \bar\Delta_f}
\end{equation}
with
\begin{equation}
\label{eq:Deltall}
{\bf \Delta_f} = \left(I^f_3-Q_f\sin^2\theta_W\right)\ \cos 2\beta M_Z^2
\end{equation}
\begin{equation}
\label{eq:Deltarr}
{\bf \bar\Delta_f} = Q_f\sin^2\theta_W \ \cos 2\beta M_Z^2
\end{equation}
and
\begin{equation}
\label{eq:mlr}
{\bf M_{LR}^2}={\bf M_{RL}^2} = 
\begin{cases}
v\left[{\bf a_f} \sin\beta -\mu {\bf Y_f} \cos\beta\right]\ , & {\tilde u}-{\rm type\ sfermion}\\
v\left[{\bf a_f} \cos\beta -\mu {\bf Y_f} \sin\beta\right]\ , & {\tilde d}-{\rm type\ sfermion}
\end{cases}\ \ \ ,
\end{equation}
where $\tan\beta=v_u/v_d$ gives the ratio of the vacuum expectation value of the two neutral Higgs fields, ${\bf Y_f}$ and ${\bf a_f}$ are the $3\times 3$ Yukawa and soft triscalar couplings and $\mu$ is the supersymmetric coupling between the two Higgs supermultiplets. The matrices ${\bf m_Q^2}$, ${\bf m_{\bar f}^2}$, and ${\bf m_q^2 }$ are the mass matrices for the LH squarks, RH squarks, and quarks, respectively. It is often customary to assume that ${\bf a_f}\propto {\bf Y_f}$, in which case one may diagonalize  ${\bf M_{LR}^2}$ by the same rotation that diagonalizes the fermion mass matrices and leads to a magnitude for L-R mixing proportional to the relevant fermion mass. In what follows, however, we will avoid making this assumption. Indeed, studies of the decay correlation parameters may provide a means of testing this alignment hypothesis.

The matrix ${\bf M_{\tilde f}^2}$ can be diagonalized by the unitary matrix ${\bf Z_f}$. The corresponding sfermion mass eigenstates ${\tilde F}_j$ are given as a linear combination of the flavor eigenstates ${\tilde f}_I$ as \begin{equation}
{\tilde F}_j = Z_f^{jI}\, {\tilde f}_I
\end{equation}
where $I=1,2,3$ indicate the flavor states ${\tilde f}_{L_I}$ and $I=4,5,6$ refer to the RH flavor states 
${\tilde f}_{R_{I-3}}$.

In general, the charginos ($\tilde\chi^+_j$) and neutralinos entering the loop graphs are mixtures of the electroweak gauginos and Higgsinos. Since the characteristics of this mixing are not crucial to our analysis and a detailed discussion can be found elsewhere ({\em e.g.}, Refs.~\cite{Rosiek:1989rs,Rosiek:1995kg}), we simply give our notation for the relevant mixing:
\be
\chi_i^0 =  N_{ij}\psi_i^0\ \ \ i,j=1 \ldots 4 \\
\ee
for the neutralinos and 
\begin{equation}
\left(\begin{array}{c}
\chi_1^+\\ \chi_2^+ \end{array}\right)
={\bf V}\left(\begin{array}{c}
\tilde{W}^+\\ \tilde{H}_u^+ \end{array}\right)\ \ \ , \ \ \ 
\left(\begin{array}{c}
\chi_1^-\\ \chi_2^- \end{array}\right)
={\bf U}\left(\begin{array}{c}
\tilde{W}^-\\ \tilde{H}_d^- \end{array}\right);\ \ \ 
\end{equation}
for the charginos. Here, the fields $\psi_i^0$ denote the $(\tilde{B}, \tilde{W}^0, \tilde{H}_d^0, \tilde{H}_u^0)$ fields. 

Using the foregoing conventions, we obtain the following contributions to the $g^S_{RR}$, $a^S_{RR}$,  and $a^{S,T}_{RL}$:
\bea
\nonumber
g^S_{RR} & = & \tilde\delta_\mu^{{\mathbf (a)}}+\tilde\delta_\mu^{{\mathbf (b.1)}}+\tilde\delta_\mu^{{\mathbf (b.2)}} \\
\label{eq:box1}
a^S_{RR} & = & \tilde\delta_\beta^{\mathbf (a)} \\
\nonumber
a^S_{RL} = -2 a^T_{RL} & = & \tilde\delta_\beta^{\mathbf (b)}
\eea
where 
\bea
\label{eq:deltamua}
\tilde\delta_\mu^{{\mathbf (a)}}&=&\frac{\alpha M_Z^2}{\pi} \; \left|U_{m1}\right|^2Z_L^{2i*}Z_L^{5i}Z_L^{1k}Z_L^{4k*}\left|N_{j1}\right|^2 \; {\mathcal F}_1\left(M_{\chi_j^0},M_{\chi_{m}^+},M_{\tilde l_i},M_{\tilde l_{k}}\right)\\
\label{eq:deltamub1}
\tilde\delta_\mu^{\mathbf (b.1)}&=&\frac{-\alpha M_Z^2}{2\pi} \; N_{j1}(N^*_{j2}-{\rm tan}\theta_W N^*_{j1}) \; N^*_{k1}(N_{k2}-{\rm tan}\theta_W N_{k1}) \\
\nonumber
& & \qquad \qquad \times \; Z_\nu^{1m*}Z_\nu^{2m}Z_L^{4i*}Z_L^{5i} \; {\mathcal F}_1\left(M_{\chi_j^0},M_{\chi_{k}^0},M_{\tilde l_i},M_{\tilde \nu_{m}}\right)\\
\label{eq:deltamub2}
\tilde\delta_\mu^{\mathbf (b.2)}&=&\frac{-\alpha M_Z^2}{2\pi} \; \; N_{j1}(N_{j2}-{\rm tan}\theta_W N_{j1}) \; N^*_{k1}(N^*_{k2}-{\rm tan}\theta_W N^*_{k1}) \\ 
\nonumber
& & \qquad \qquad \times \; Z_\nu^{1m*}Z_\nu^{2m} Z_L^{4i*} Z_L^{5i}
\; M_{\chi_j^0} \: M_{\chi_{k}^0} \; {\mathcal F}_2\left(M_{\chi_j^0},M_{\chi_{k}^0},M_{\tilde l_i},M_{\tilde \nu_{m}}\right)\\
\label{eq:deltabetaa}
\tilde\delta_\beta^{\mathbf (a)}&=&\frac{\alpha M_Z^2 V_{ud}}{3\pi} \left|U_{k1}\right|^2Z_D^{1i*}Z_D^{4i}Z_L^{1m}Z_L^{4m*}\left|N_{j1}\right|^2{\mathcal F}_1\left(M_{\chi_j^0},M_{\chi_{k}^+},M_{\tilde d_i},M_{\tilde l_{m}}\right)\\
\label{eq:deltabetab}
\tilde\delta_\beta^{\mathbf (b)}&=&\frac{-\alpha M_Z^2 V_{ud}}{3\pi} U_{j1}V_{j1}^* Z_U^{1i*}Z_U^{4i}Z_L^{1m}Z_L^{4m *}\left|N_{k1}\right|^2M_{\chi_j^+}M_{\chi_{k}^0}{\mathcal F}_2\left(M_{\chi_j^+},M_{\chi_{k}^0},M_{\tilde u_i},M_{\tilde l_{m}}\right)
\eea
and where we have defined the loop functions 
\begin{equation}\label{eq:f}
{\mathcal F}_{n}\left(m_a,m_b,m_c,m_d\right)\equiv\int_0^1{\rm d}x\int_0^{1-x}{\rm d}y\int_0^{1-x-y}{\rm d}z\left[x\ m_a^2+y\ m_b^2+z\ m_c^2+(1-x-y-z)m_d^2\right]^{-n} \ \ \ .
\end{equation}

\section{Phenomenological Constraints and Implications}
\label{sec:constraints}

We now analyze the possible magnitude of the box graph contributions.  At first glance, the results in Eqs. (\ref{eq:deltamua}-\ref{eq:deltabetab}) exhibit the expected scaling with masses and couplings: ${\tilde\delta}\sim (\alpha/4\pi)\times (M_Z/{\tilde M})^2$, where ${\tilde M}$ is a generic superpartner mass. Thus, one expects these contributions to be of order $10^{-3}$ when ${\tilde M}$ is comparable to the electroweak scale. However, the prefactors involving products of the sfermion mixing matrices can lead to substantial departures from these expectations. (The impact of the neutralino and chargino mixing matrices $N_{jk}$, $U_{jk}$ {\em etc.} is less pronounced). In particular, there are two general classes box graph contributions: those which depend on slepton flavor mixing ($\delta_\mu^{(b.1,2)}$) and those which depend on L-R mixing ($\delta_\mu^{(a)}$ and $\delta_\beta^{(a,b)}$). We examine each class of contributions independently by performing a numerical scan over MSSM parameter space while taking into consideration the results of direct searches for superpartners, precision electroweak data, and LFV studies. In particular,  
we attempt to constrain the viable parameter space -- and thus the magnitude of these box graph contributions -- by requiring consistency with the experimental bounds for the branching ratio for $\mu \to e\gamma$ and for $(g_\mu-2)$.

\subsection{Lepton Flavor Mixing Contributions}

The contributions to $g^S_{RR}$ from $\delta_\mu^{(b.1,2)}$ depend on
the products of stermion mixing matrices $Z_\nu^{1m*}Z_\nu^{2m}$ and $Z_L^{5i}Z_L^{4i*}$,
which are non-vanishing only in the presence of flavor mixing among the first two generations of sneutrinos and RH charged sleptons, respectively. The existence of such flavor mixing also gives rise to lepton flavor violating (LFV) processes such as $\mu\to e\gamma$ and $\mu\to e$ conversion and, indeed, the  products $Z_\nu^{1m*}Z_\nu^{2m}$ and $Z_L^{5i}Z_L^{4i*}$ enter the rates for such processes at one-loop order. Consequently, the non-observation of LFV processes leads to stringent constraints on these products of mixing matrix elements.  

To estimate the order of magnitude for these constraints, we focus on the rate for the decay $\mu\to e\gamma$, which turns out to be particularly stringent. In principle, one could also analyze the constraints implied by limits on the $\mu\to e$ conversion and $\mu\to 3e$ branching ratios. This would possibly make our conclusions on the maximal size of the flavor violating contributions to $g_{RR}^S$ more severe, but it would not change our main conclusion: lepton flavor mixing contributions to $g_{RR}^S$ are unobservably small. 

Experimentally, the most stringent bound on the corresponding branching ratio  has been obtained by the MEGA collaboration\cite{Brooks:1999pu}:
\be
\label{eq:mega}
Br(\mu\to e\gamma) \equiv \frac{\Gamma(\mu^+\to e^+\gamma)}{\Gamma(\mu^+\to e^+\nu{\bar\nu})}
< 1.2 \times 10^{-11}\qquad\qquad {\rm 90\% \; C.L.}  \ \ \ 
\ee
Theoretically, a general analysis in terms of slepton and sneutrino mixing matrices has been given in Ref.~\cite{lfv}. Using the notation of that work, we consider those contributions to the $\mu\to e\gamma$ amplitude that contain the same combinations for LFV mixing matrices as appear in the $\tilde\delta_\mu^{(b)}$. For simplicity, we also set (in the present analytical estimate, but not in the following numerical computation) the chargino and neutralino mixing matrices to unity and neglect contributions that are suppressed by factors of $m_\mu/M_W$. With these approximations, the combination $Z_\nu^{1m}Z_\nu^{2m}$ appears only in the first term of the chargino loop amplitude $A^{(c)R}_2$ according to the notation of Ref.~\cite{lfv}. Setting the chargino mixing to 1 (or, equivalently, considering the pure wino contribution alone), gives
\begin{equation}
A^{(c)R}_2\simeq\frac{\alpha}{8\pi\sin^2\theta_W}\frac{1}{m_{\widetilde \nu_m}^2}\left(Z_\nu^{1m}Z_\nu^{2m}\right)f^{(c)}(x_m)+\cdots\ ,\quad x_m=\left(\frac{m_{\widetilde \nu_m}}{m_{\widetilde W}}\right)
\end{equation}
where $f^{(c)}(x)$ is a loop function. 
Analogously, the combination $Z_L^{4i}Z_L^{5i}$ appears only in the first term of the neutralino loop amplitude $A^{(n)L}_2$. This time the amplitude reads (again considering only the pure {\em bino} loop)
\begin{equation}
A^{(n)L}_2\simeq\frac{\alpha}{4\pi\cos^2\theta_W}\frac{1}{m_{\widetilde L_i}^2}\left(Z_L^{4i}Z_L^{5i}\right)f^{(n)}(x_i)+\cdots\ ,\quad x_i=\left(\frac{m_{\widetilde L_i}}{m_{\widetilde B}}\right)
\end{equation}
The resulting muon decay widths respectively read
\begin{equation}
\Gamma^{(c)}(\mu\rightarrow e\gamma)\simeq\frac{\alpha}{4}\frac{\alpha^2}{(8\pi\sin^2\theta_W)^2}\frac{m_\mu^5}{m_{\widetilde\nu_m}^4}\left(Z_\nu^{1m}Z_\nu^{2m}\right)^2\left(f^{(c)}(x_m)\right)^2
\end{equation}
and
\begin{equation}
\Gamma^{(n)}(\mu\rightarrow e\gamma)\simeq\frac{\alpha}{4}\frac{\alpha^2}{(4\pi\cos^2\theta_W)^2}\frac{m_\mu^5}{m_{\widetilde L_i}^4}\left(Z_L^{4i}Z_L^{5i}\right)^2\left(f^{(n)}(x_i)\right)^2
\end{equation}

For simplicity, we consider two extremes: $m_{\widetilde\nu_m}\approx m_{\widetilde L_i}\approx m_{\widetilde W}\approx m_{\widetilde B}$=100, 1000 GeV $\equiv{\tilde M}$. For either choice, we find
\begin{equation}
\left(f^{(c)}(1)\right)^2\simeq\left(f^{(n)}(1)\right)^2\simeq0.007
\end{equation}
Inserting the numerical values, and requiring that $\Gamma^{(n,c)}\lesssim 10^{-30}$ GeV as required by the limit (\ref{eq:mega}), we find that 
\begin{equation}
\left(Z_L^{4i}Z_L^{5i}\right)_{\rm max}\approx \left(Z_\nu^{1m}Z_\nu^{2m}\right)_{\rm max}<
\begin{cases}
10^{-3} & {\rm for} \ {\tilde M} =
100\ {\rm GeV}\\
10^{-1} & {\rm for} \ {\tilde M} =1000\ {\rm GeV}
\end{cases}
\end{equation}
This implies that for superpartner masses of the order of 100 GeV, the amplitudes ${\tilde\delta}_\mu^{(b.1,b.2)}$ are suppressed by a factor $10^{-6}$ relative to the naive expectations discussed above, while for 1000 GeV masses by a factor $10^{-2}$. In this latter case, however, the loop functions in the amplitudes ${\tilde\delta}_\mu^{(b)}$ experience a further suppression factor of order $10^{-2}$. Thus, the magnitude of the ${\tilde\delta}_\mu^{(b)}$ should be no larger than $\sim 10^{-7}$. 

\begin{table}[!b]
\begin{center}
\begin{tabular}{|c|c|c|c|c|}\hline
$\mu$ & $m_1$ & $m_2$ & $({\bf M^2_{LL} })_{ij}, ({\bf M^2_{RR} })_{ij}$  & $\tan\beta$\\
\hline
$30\div10000$ &  $2\div1000$ &  $50\div1000$ &  $10^2\div2000^2$ &  $1\div60$\\
\hline
\end{tabular}
\end{center}
\caption{\it\small Ranges of the MSSM parameters used to generate
the models shown in Fig.~\protect{\ref{fig:lfv}}. Here, $\mu$ is the usual higgsino mass term, while $m_{1,2}$ indicate the soft supersymmetry breaking U$(1)_Y$ and SU(2) gaugino masses. The matrices
${\bf M^2_{LL} }$ and ${\bf M^2_{RR} }$ are symmetric; hence, we scanned over 6 independent masses within the specified range.
All masses are in GeV.} \label{tab:scan2}
\end{table}

\begin{figure}[!t]
\begin{center}
\mbox{\hspace*{-0.7cm}\epsfig{file=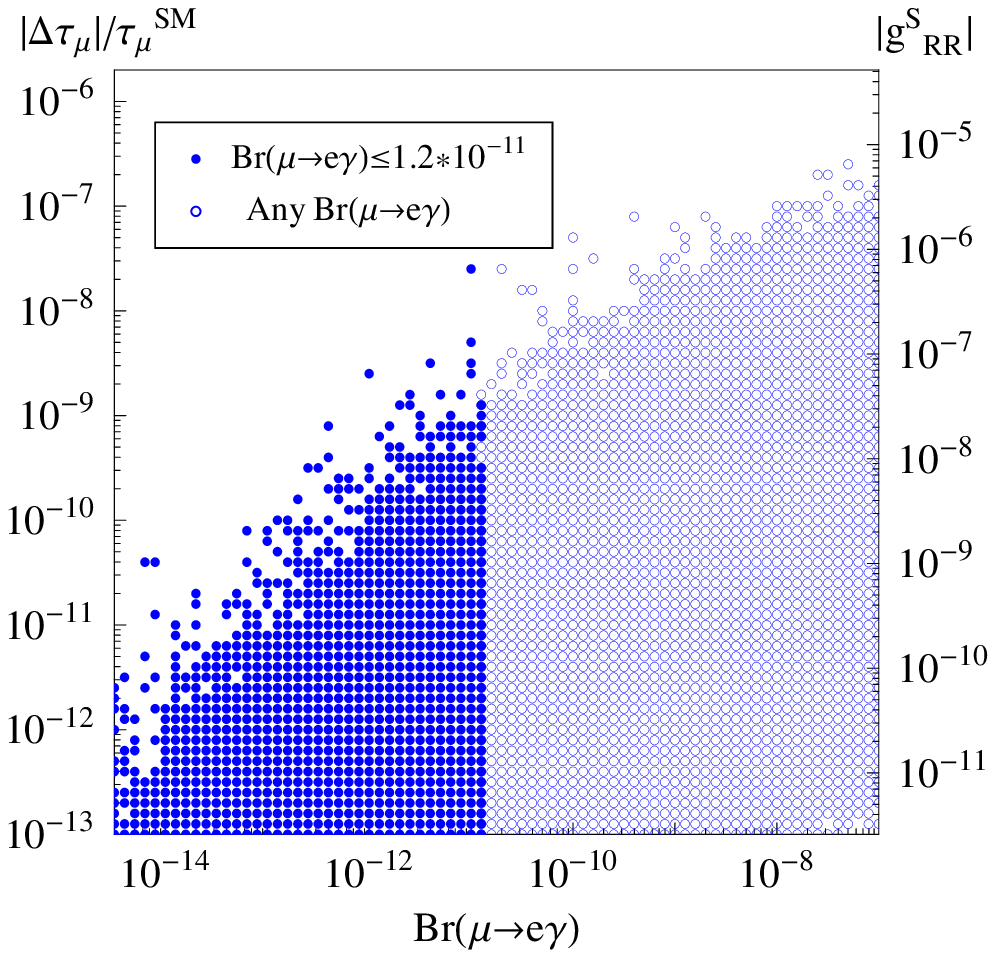,height=9.2cm}\qquad \epsfig{file=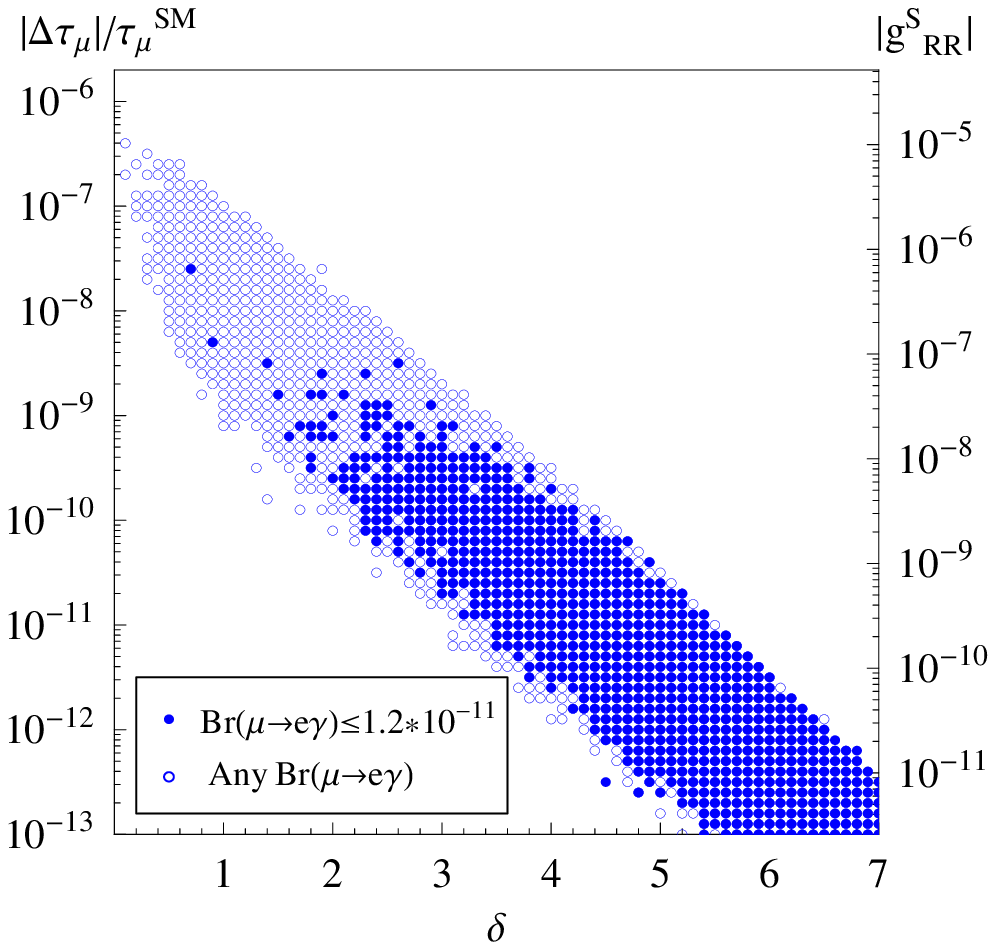,height=9.2cm}}\\
\hspace*{1cm}(a)\hspace*{10cm}(b)
\end{center}
\caption{\it\small 
A scatter plot showing $|\Delta\tau_\mu|/\tau_\mu^{\rm SM}$ (left vertical axis) and $|g^S_{\rm RR}|$ (right vertical axis), as functions of Br($\mu\rightarrow e\gamma$) (a) and of $\delta_{\rm LFV}$ (b) (smaller $\delta_{\rm LFV}$ means more lepton flavor mixing; see the text for the precise definition). Filled circles represent models consistent with the current bound Br($\mu\rightarrow e\gamma$) $\le 1.2\times 10^{-11}$, while empty circles denote all other models.}
\label{fig:lfv}
\end{figure}

We substantiate the previous estimates by performing a numerical scan over the parameter space of the $CP$-conserving MSSM \cite{Chung:2003fi}. We do not implement any  universality assumption in the slepton soft supersymmetry breaking mass sector or in the gaugino mass terms.  However, in this section only, we neglect L-R mixing and consider flavor mixing between the first and second generation sleptons only.  
Under these assumptions, the mixing that causes non-vanishing $\tilde{\delta}_\mu^{(b.1,2)}$ stems solely from off-diagonal elements in the two 2$\times$2 slepton mass matrices ${ \bf M^2_{LL}}$ and ${ \bf M^2_{RR}}$.  
We scan independently over all the parameters indicated in Table~\ref{tab:scan2}, within the specified ranges. For all models, we impose constraints from direct supersymmetric-particles searches at accelerators and require the lightest supersymmetric particle (LSP) to be the lightest neutralino (see also \cite{Profumo:2004at} for more details).

The result of this scan is shown in Fig.~\ref{fig:lfv}.  
Although general models can accommodate $g^S_{RR} \sim 10^{-5}$, the current constraint on Br($\mu\rightarrow e\gamma$) severely restricts the available parameter space and reduces the allowed upper limit on $g^S_{RR}$ by over an order of magnitude, as shown in Fig.~\ref{fig:lfv} (a).  
It is also instructive to exhibit the sensitivity of $g^S_{RR}$ to the degree of flavor-mixing and to show the corresponding impact of the LFV searches. To that end, we quantify the amount of lepton flavor mixing by a parameter $\delta_{\rm LFV}$, defined as   
\be
\delta_{\rm LFV} = |\delta_L| + |\delta_R| \;,
\ee
where
\be
\delta_{L} = \log \left( \frac{2 \: ({ \bf M^2_{LL}})_{12} }{({\bf M^2_{LL}})_{11}+({\bf M^2_{LL} })_{22}} \right) 
\qquad \qquad
\delta_{R} = \log \left( \frac{2 \: ({ \bf M^2_{RR}})_{12} }{({\bf M^2_{RR}})_{11}+({\bf M^2_{RR} })_{22}} \right) 
\ee
and {\em e.g.},  $({\bf M^2_{LL}})_{ij}$ is the $(i,j)$-th component of left-handed slepton mass matrix. 
For example, if $|\delta_L|$ is close to zero, then there is a large flavor mixing contribution from left-handed sleptons; but if $|\delta_L|$ is large, then this flavor mixing is suppressed.
Since the amplitudes $\tilde{\delta}_\mu^{(b.1,2)}$ depend on flavor mixing among {\it{both}} LH and RH sleptons, they contribute only if both $\delta_L$ and $\delta_R$ are small.  (In contrast, Br($\mu\rightarrow e\gamma$) survives in the presence of flavor mixing among {\it{either}} LH or RH sleptons.)  Naturally, the flavor mixing contribution to $g^S_{RR}$ is largest when $\delta_{LFV}$ is smallest, as shown in Fig.~\ref{fig:lfv}(b). We note that to obtain a large flavor mixing contribution to $g^S_{RR}$, it is not sufficient simply to have maximal mixing (i.e. $|Z_L^{ij}| = 1/\sqrt{2}$); in addition, one needs the absence of a degeneracy among the slepton mass eigenstates, or else the sum over mass eigenstates [{\em e.g.},  sum over $i$ and $i^\prime$ in Eqn~(\ref{eq:deltamub1})] will cancel.  In any case, we conclude that the flavor mixing box graph contributions are too small to be important  for the interpretation of the next generation muon decay experiments, where the precision in $\tau_\mu$ is expected to be of the order of one ppm.

\subsection{Left-Right Mixing Contributions}


Significantly larger contributions to $g^S_{RR}$ can arise from ${\tilde\delta}_\mu^{(a)}$, which requires only L-R mixing among same generation sleptons. As noted earlier, L-R mixing can be appreciable when either $Y_f\, |\mu |$ or $a_f$ is of order the electroweak or soft breaking scale. For the first and second generation sfermions, taking $Y_f\, |\mu|\sim v$ implies that $|\mu|$ lies many orders of magnitude above the electroweak scale. On general grounds, one expects $|\mu | >> v$, thereby generically requiring some degree of fine-tuning to obtain proper EWSB. The need for such fine tuning  is sometimes dubbed as ``the $\mu$ problem''. However, the relevance of fine-tuning-based arguments has been recently questioned, based for instance upon the evidence for a non-vanishing extremely small value for the cosmological constant, suggesting that nature does not care about theoretical prejudices for ``naturalness''. Thus, it is interesting to explore the phenomenological consequences of large values of $|\mu|$, which  would include  large L-R mixing\footnote{The largest non-$(V-A)\otimes(V-A)$ induced by the loop effects studied here arise when the virtual charginos and neutralinos are SU(2$)_L$ gaugino like -- rather than bino or higgsino like -- and have  masses close to the  SU(2$)_L$ gaugino mass $m_2$. Thus, large values of $|\mu|$ would not suppress the one-loop contributions.}. Large values of $\mu$ could in principle lead to tachyonic masses for those scalar fermions featuring large $Y_f$, such as the stop, the sbottom and the stau. However, since at this point we refrain from any assumption on the structure of the soft supersymmetry breaking masses, in principle the mass spectrum of third generation sfermions can be heavier than that of the first two generations, without violating any phenomenological constraint.


In the case of large values for $a_f$, the possible existence of nearly flat directions in the MSSM potential leads to additional constraints on the size of the scalar trilinear couplings from the condition of avoiding charge and color breaking minima. Quantitatively, one can express those constraints in the form \cite{strumia}
\begin{eqnarray}
\nonumber {\bf a^2_u}&\lesssim&3\ {\bf Y^2_u}\left(\mu^2_u+{\bf m^2_{\tilde Q}}+{\bf m^2_{\tilde u}}\right)\\
\nonumber {\bf a^2_d}&\lesssim&3\ {\bf Y^2_d}\left(\mu^2_d+{\bf m^2_{\tilde Q}}+{\bf m^2_{\tilde d}}\right)\\
 {\bf a^2_e}&\lesssim&3\ {\bf Y^2_e}\left(\mu^2_d+{\bf m^2_{\tilde L}}+{\bf m^2_{\tilde e}}\right)\label{eq:chargecolor}
\end{eqnarray}
where $\mu^2_{u,d}\equiv m^2_{h_{u,d}}+|\mu|^2$. The conditions above can be fulfilled for large values of ${\bf a_f}$ for accordingly large values of ${\bf Y_f}\times {\bf m_{\tilde f}}$
or ${\bf Y_f}\times {\bf \mu_{u,d}}$. Since  $g^S_{RR}$ depends on the product $\mathcal F_1\times|(Z_L^{22})^*Z_L^{52}|$, the presence of large scalar fermion masses leads to a suppression of $g^S_{RR}$ via the loop functions $\mathcal F_1$. Thus, large values of $g^S_{RR}$ are possible only when L-R mixing is nearly maximal and the constraints of Eqs.~(\ref{eq:chargecolor}) are satisfied with large values of $\mu^2_{u,d}$. Making use of the EWSB conditions, one can express $\mu^2_{u,d}$ as functions of $M_Z$, $|\mu|$ and of the CP-odd Higgs mass, $m_A$. One may then achieve arbitrarily large values for the right hand sides of Eqs.~(\ref{eq:chargecolor}), and hence of the trilinear scalar couplings, as long as, for instance, $m_A$ is sufficiently large, independently of the other sfermion soft supersymmetry breaking mass terms. In this case, all but the lightest CP-even Higgs scalar would be light enough to be observable in future collider studies at the LHC or a 1 TeV $e^+ e^-$ linear collider, a situation that is not ruled out by present Higgs searches\footnote{The heavy Higgs sector does not enter in the loops contributing to any of the quantities at stake here, and the size of $m_A$ therefore does not affect our results.}. Otherwise, one can fulfill Eq.~(\ref{eq:chargecolor}) above assuming large values for $|\mu|$. 

The possibility that one or more of the new scalar particles in the MSSM is considerably heavier than the electroweak scale is not a novel idea. Recently,  models have been proposed that feature a ``split'' supersymmetry, wherein the soft breaking scalar and gaugino masses lie at hugely separated scales \cite{Arkani-Hamed:2004yi} with the gaugino masses remaining near the electroweak scale. There is no  {\em a priori} reason not to consider an analogous situation in which the split sector is the Higgs sector alone, instead of the whole scalar sector. Even though in particular models -- such as minimal supergravity, wherein the sfermion and the heavy Higgs sector are connected at the GUT scale and  L-R mixing in the first two generations must be suppressed -- the possibility of having large L-R mixing and a split Higgs sector is both viable and motivated in other supersymmetric scenarios. For example, a split scalar soft breaking mass sector can arise straightforwardly  in the context of supersymmetric grand unification \cite{deBoer:1994dg,Baer:2000gf}. Specifically, in supersymmetric SO(10) grand unified theories -- which are particularly interesting in light of  recent data on neutrino mixing and mass splittings  -- the matter multiplets of each generation belong to the 16 dimensional spinorial representation of SO(10), while the Higgs multiplets inhabit a single 10 dimensional fundamental representation. The corresponding soft breaking mass scales are completely unrelated  and a substantial hierarchy between the two is not unreasonable \cite{Baer:2000gf,nuhm}. 

In view of these considerations, we turn to a detailed phenomenological analysis of large L-R mixing among first and second generation sfermions. In the case of smuon L-R mixing, some considerations follow from the present value for the muon anomalous magnetic moment, or $(g_\mu-2)$, which is a chirality odd operator and which can arise from L-R mixing in one-loop graphs \cite{Martin:2001st}. The only supersymmetric contribution $\delta a_\mu$ to $a_\mu\equiv (g_\mu-2)/2$ proportional to one single power of the ratio of the muon mass and of supersymmetric particles is in fact proportional to the smuon L-R mixing, and reads
\be\label{eq:gm2}
\delta a^{\rm LR-mix}_\mu=\frac{m_\mu}{16\pi^2}\sum_{i,m}\frac{m_{\chi^0_i}}{3m_{\tilde\mu_m}^2}{\rm Re}[g_1N_{i1}(g_2N_{i2}+g_1N_{i1})Z_L^{2m*}Z_L^{5m}]F^N_2(m_{\chi^0_i}^2/m_{\tilde\mu_m}^2),
\ee
where $F^N_2$ is the appropriate loop function specified in \cite{Martin:2001st}. The expression features the same dependence upon the smuon mixing matrix as does Eq.~(\ref{eq:deltamua}). Under the widely considered alignment and electroweak scale $\mu$ assumptions, this term is usually suppressed, as the smuon L-R mixing is also suppressed. Here, however, we drop that hypothesis, and allow for large L-R mixing: this, in general, enhances the aforementioned contribution, and we therefore expect that the experimental constraints on $(g_\mu-2)$ will set limits on $g^S_{RR}$. In particular, assuming a common mass $\tilde M$ for all the supersymmetric particles, and maximal L-R mixing, Eq.~(\ref{eq:gm2}) approximately reduces to
\be
\delta a^{\rm LR-mix}_\mu\approx\frac{g_1^2}{4\pi}\frac{1}{12\pi}\left(\frac{m_\mu}{\tilde M}\right)\approx10^{-7}\quad {\rm for}\ \tilde M\sim100\ {\rm GeV},
\ee
where $g_1$ indicates the $U(1)_Y$ gauge coupling. We consider here the 95\% C.L. limit on beyond-the-SM contributions to $(g_\mu-2)$ as quoted in Ref.~\cite{gm2},
\be
a_\mu^{\rm exp}-a_\mu^{\rm th-SM}=(25.2\pm9.2)\times10^{-10},
\ee
bearing in mind that a more conservative approach to the evaluation of the sources of theoretical uncertainty in the SM contribution could inflate the error associated with $a_\mu^{\rm th-SM}$ (see {\em e.g.} Ref.~\cite{mjrmwise} for the hadronic light-by-light contribution).

\begin{figure}[!t]
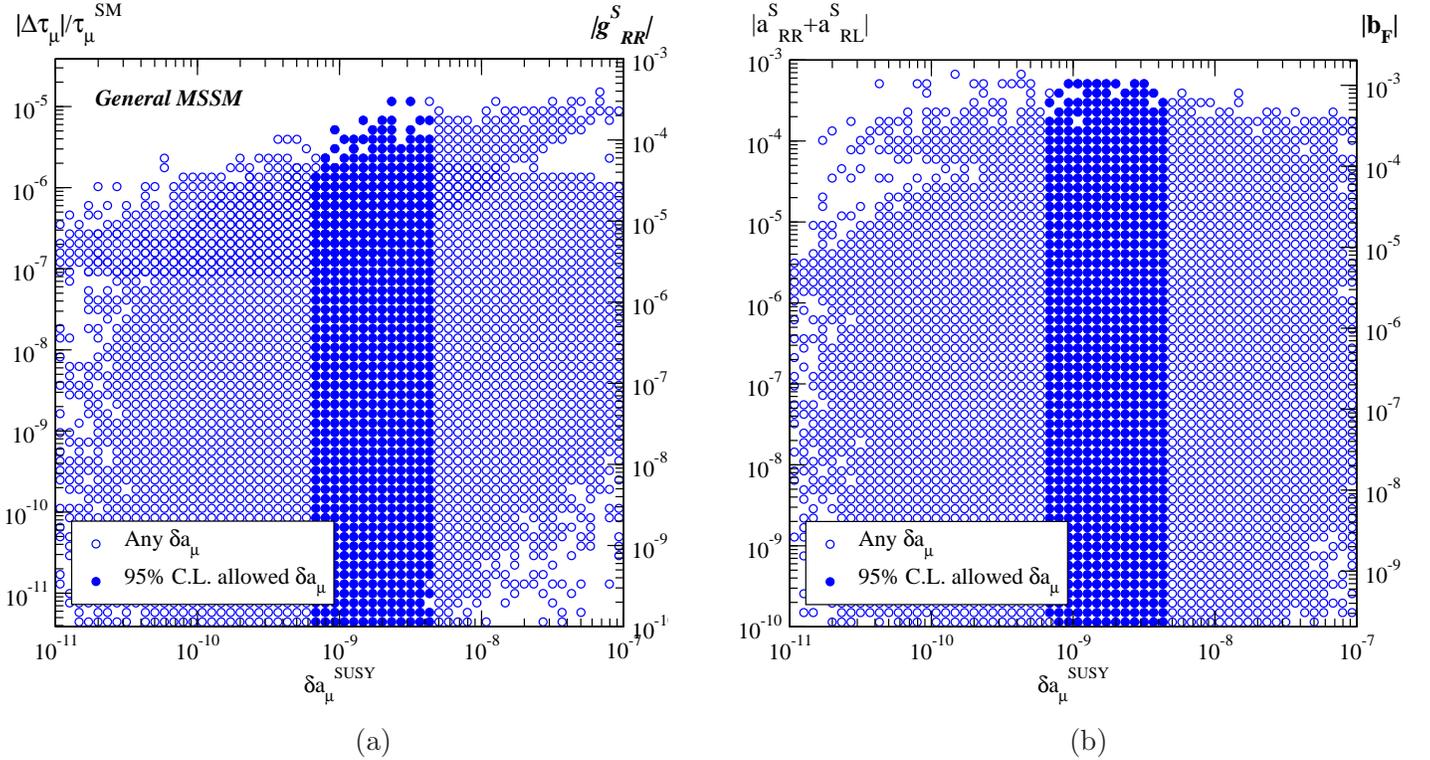

\begin{center}
\mbox{\hspace*{-0.7cm}\epsfig{file=amu_mu.eps,height=9.4cm}\qquad \epsfig{file=amu_beta.eps,height=9.4cm}}\\
\hspace*{1cm}(a)\hspace*{9cm}(b)
\end{center}
\caption{\it\small 
A scatter plot showing $|\Delta\tau_\mu|/\tau_\mu^{\rm SM}$ (a) and $|g^S_{\rm RR}|$ (b), relative to muon decay, left, and $a^S_{\rm RR}+a^S_{\rm RL}$, relative to $\beta$ decay, right, as a function of the supersymmetric contribution to the muon anomalous magnetic moment $\delta a_\mu$. Filled circles represent models consistent with the current 95\% C.L. range for beyond the standard model contributions to $(g_\mu-2)$, while empty circles denote all other models.}
\label{fig:amu}
\end{figure}

A large value for $\delta a_\mu$, however, does not imply automatically a large $g^S_{RR}$, since the latter also depends upon the mixing in the selectron sector, to which $\delta a_\mu$ is blind. On the other hand, large values of $|g^S_{RR}|$ in general should produce a sizable $\delta a_\mu$, although the possibility of cancellations with other terms, and the different loop function structures can lead, in principle, to a suppression of $\delta a_\mu$ even for large $g^S_{RR}$. This possibility is illustrated in Fig.~\ref{fig:amu}, showing a scan over MSSM parameters (discussed below). As indicated by the results of this scan, imposing the $(g_\mu-2)$ constraints restricts, but does not exclude, the possibility of obtaining relatively large values of $|g^S_{RR}|$.

Large L-R mixing in the selectron sector also induces potentially large contributions to the electron $(g_e-2)$. We did take into account this further constraint, although we generically find that it gives less severe bounds on the models we consider than those from $(g_\mu-2)$.


In contrast to the situation for $\mu$-decay, the box graph contributions to the semileptonic parameters $a^S_{RR}$ and $a^{S,T}_{RL}$ live entirely on L-R mixing among first generation sleptons and squarks. To our knowledge, there exist no strong bounds on such mixing from precision electroweak measurements or searches for rare or SM-forbidden processes. Consequently, we will consider the possibility of maximal L-R mixing which simply requires that 
$|M_{LR}^2| \sim |M_{LL}^2-M_{RR}^2|$. From Eqs. (\ref{eq:mll}-\ref{eq:mlr}), this situation amounts to having $|M_{LR}|$ of order the electroweak scale and $m_F^2$ not too different from $m_{\bar f}^2$. The foregoing discussion of charge and color breaking minima applies to this case as well as to $\mu$-decay. 

\begin{table}[!b]
\begin{center}
\begin{tabular}{|c|c|c|c|c|c|c|}\hline
$|\mu|$ & $|m_1|$ & $|m_2|$ & $|m_3|$ & $m_{\widetilde F}$ & $a_F$ & $\tan\beta$\\
\hline
$30\div10000$ &  $2\div1000$ &  $50\div1000$ &  $m_{\rm\sss LSP}\div10000$ & $(1\div10)m_{\rm\sss LSP}$ &  $\pm(m^2_{\widetilde F}/v)$ &  $1\div60$\\
\hline
\end{tabular}
\end{center}
\caption{\it\small Ranges of the MSSM parameters used to generate
the models shown in Figs.~\protect{\ref{fig:amu}} and
\protect{\ref{fig:corr}}. All masses are in GeV, and
$m_{\rm\sss LSP}\equiv{\rm min}(|\mu|,|m_1|,|m_2|)$. $m_3$ indicates the gluino mass and the quantity
$m_{\widetilde F}$
 the various soft supersymmetry breaking masses, which we independently sampled; we dub the mass of the corresponding SM fermion $F$ as $m_F$.}\label{tab:scan}
\end{table}
Taking into account the foregoing considerations, we carry out a numerical analysis of the magnitude of the SUSY contributions to $g^S_{RR}$, $a^S_{RR}$, and $a^{S,T}_{RL}$. 
As before, we consider the  $CP$-conserving MSSM  \cite{Chung:2003fi}, and proceed to a random scan over its parameter space. We do not resort to any universality assumption, neither in the scalar soft supersymmetry breaking mass sector, nor in the gaugino mass terms nor in the soft-breaking trilinear scalar coupling sector, and scan independently over all the parameters indicated in Table~\ref{tab:scan}, within the specified ranges. We indicate with $m_{\widetilde F}$ a generic (left or right, that we independently sample) scalar fermion soft supersymmetry breaking mass (corresponding to a standard model fermion whose mass is $m_F$), and with $m_{\rm\sss LSP}$ the smallest mass parameter entering the neutralino mass matrix (namely, $m_1,m_2$ and $\mu$), in absolute value. The trilinear scalar couplings $a_F$, corresponding to the fermion $F$, are sampled in the range $-m_{\widetilde F,{\rm L}}m_{\widetilde F,{\rm R}}/v<a_F<m_{\widetilde F,{\rm L}}m_{\widetilde F,{\rm R}}/v$. For all models, we impose constraints from direct supersymmetric-particles searches at accelerators, rare processes with a sizable potential supersymmetric contribution, the lower bound on the mass of the lightest $CP$-even Higgs boson, and precision electroweak tests. We also require the lightest supersymmetric particle (LSP) to be the lightest neutralino (see also \cite{Profumo:2004at} for more details) and avoid parameter choices that lead to tachyonic solutions. In our scan, for a given parameter space point, we set the scale $m_A$ of the heavy Higgs sector to the value saturating the color and charge breaking minima constraints in Eq.(\ref{eq:chargecolor}). 

When the trilinear scalar couplings are maximal ({\em i.e.} close to the boundary of the sampling region), the fulfillment of Eq.(\ref{eq:chargecolor}) for the selectrons forces us to consider relatively large values of $m_A\simeq m^2_{\widetilde e}/m_e$. As stressed above, such large values are theoretically and phenomenologically acceptable, and do not affect in any way our results. The resulting contributions to the lightest $CP$-even Higgs mass, namely
\begin{equation}
\Delta m^2_h\simeq \frac{a^2_f}{16 \pi^2}N_c\ln\left(\frac{m_{\widetilde f}^2}{m_h^2}\right)
\end{equation}
are also generically small and compatible with the experimental lower bound on $m_h$.
Another significant effect driven by large soft supersymmetry breaking scalar masses in the Higgs sector is a mis-match between the D-term couplings and the usual weak scale gauge couplings, giving rise to potentially large logarithmic corrections to the lightest Higgs mass. These were implicitely taken into account in our numerical evaluation of $m_h$, that makes use of an effective supersymmetric mass scale that depends upon the values of the soft supersymmetry breaking parameters.

\begin{figure}[!t]
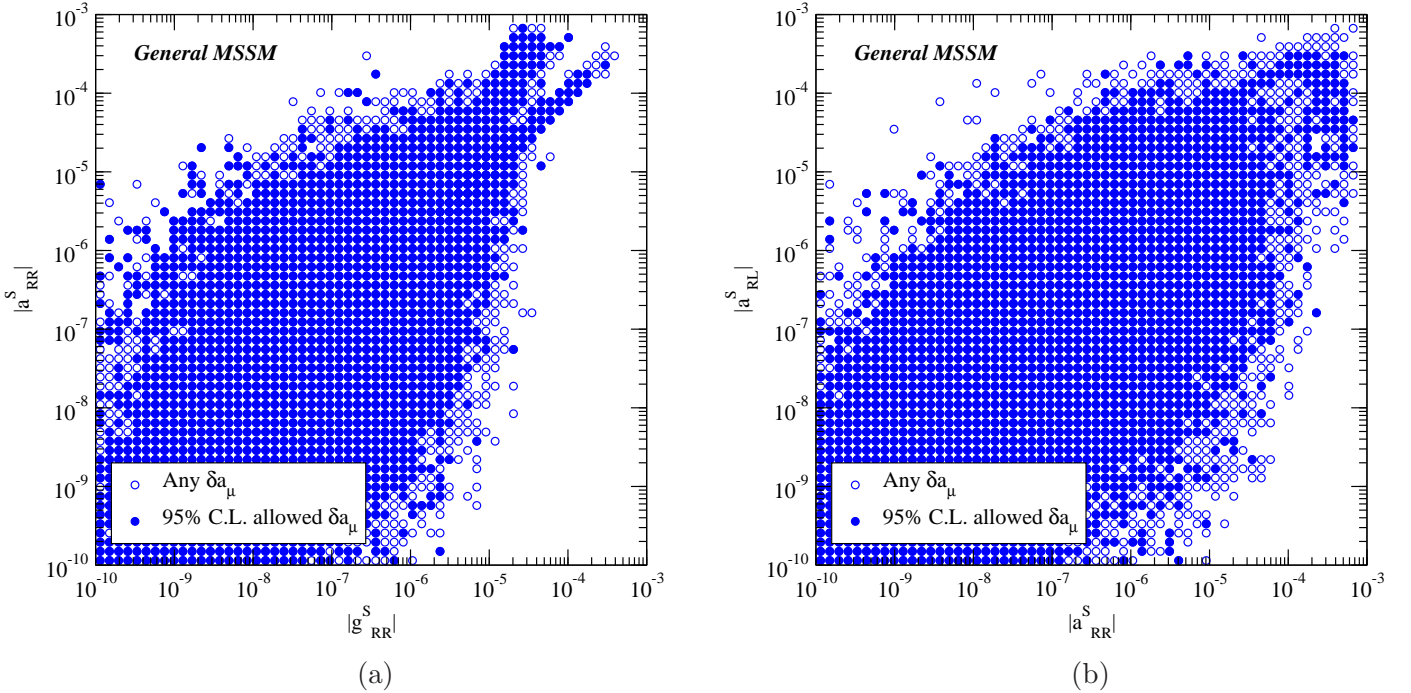

\begin{center}
\mbox{\hspace*{-0.7cm}\epsfig{file=mu_beta.eps,height=8.5cm}\qquad \epsfig{file=beta_beta.eps,height=8.5cm}}\\
\hspace*{1cm}(a)\hspace*{9cm}(b)
\end{center}
\caption{\it\small 
The correlation between $a^S_{\rm RR}$ and $g^S_{\rm RR}$ (a) and between $a^S_{\rm RR}$ and $a^S_{\rm RL}$ (b). Filled circles represent models consistent with the current 95\% C.L. range for beyond the standard model contributions to $(g_\mu-2)$, while empty circles denote all other models.}
\label{fig:corr}
\end{figure}
\begin{figure}[!t]
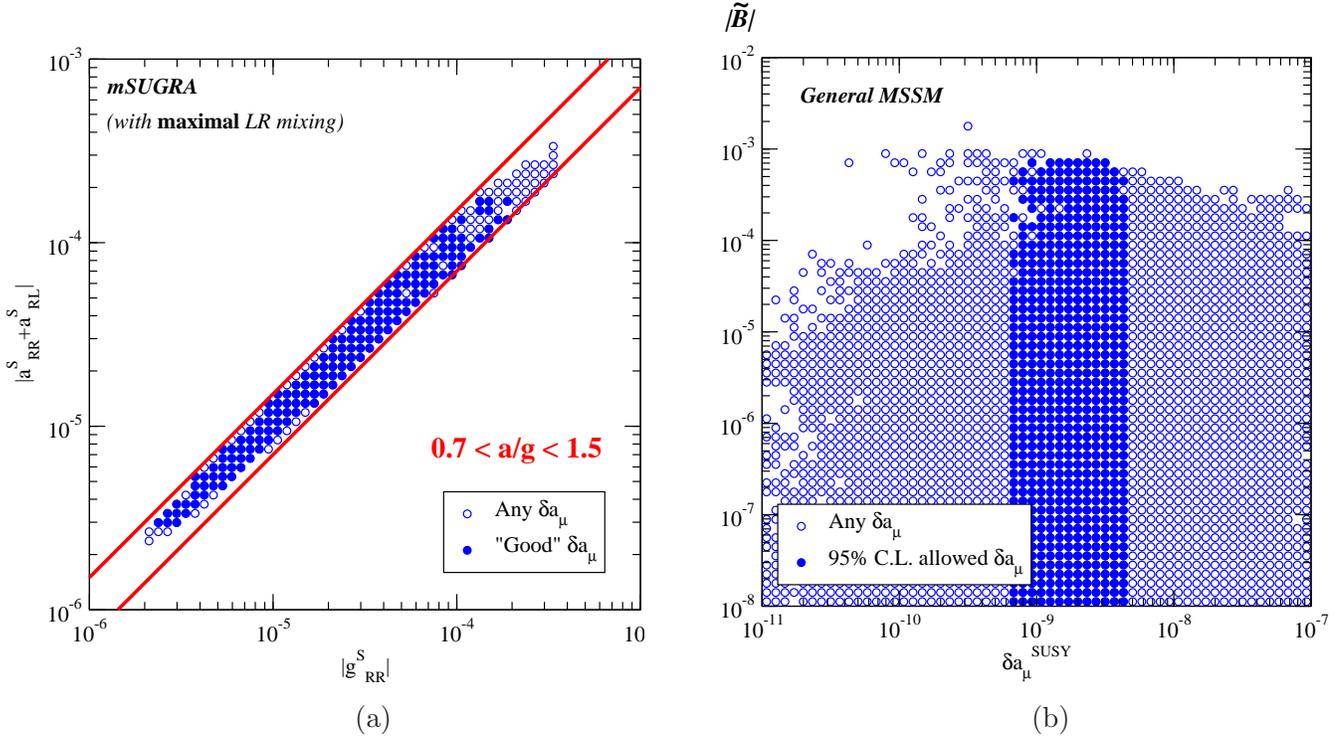

\begin{center}
\mbox{\hspace*{-0.7cm}\epsfig{file=enh_g_a.eps,height=8.5cm}\qquad \epsfig{file=tildeb.eps,height=9.1cm}}\\
\hspace*{0.5cm}(a)\hspace*{8.5cm}(b)
\end{center}
\caption{\it\small 
{\em (a)}: The correlation between $a^S_{\rm RR}+a^S_{\rm RL}$ and $g^S_{\rm RR}$ in models with a ``{\em minimal supergravity}'' slepton and squark soft supersymmetry breaking mass pattern (i.e. with correlated slepton and squark masses) and maximal left-right mixing (i.e., where the trilinear scalar couplings have been set to the values corresponding to a maximal contribution to the quantities of interest). The scalar mass universality condition at the GUT scale dictates the strong correlation between the two quantities $0.7\lesssim (a^S_{\rm RR}+a^S_{\rm RL})/g^S_{\rm RR}\lesssim 1.5$. Again, filled circles represent models consistent with the current 95\% C.L. range for beyond the standard model contributions to $(g_\mu-2)$, while empty circles denote all other models. {\em (b)}: MSSM-induced non-$(V-A)\otimes(V-A)$ contributions to the energy-dependence of the $\beta$-decay neutrino asymmetry parameter, $B$. Here, we have scaled out the energy-dependence and have plotted ${\tilde B} = B/(\Gamma m/E)$ for various randomly generated MSSM parameters. As before, dark circles indicate models consistent with $(g_\mu-2)$ We have also assumed $g_S/g_V=1=g_T/g_A$}
\label{fig:enh}
\end{figure}

We show the results of our scan in Fig.~\ref{fig:amu}-\ref{fig:enh}. In particular, we indicate in Fig.~\ref{fig:amu}, (a), the values of $|\Delta\tau_\mu|/\tau_\mu^{\rm SM}$ (left axis) and $|g^S_{\rm RR}|$ (right axis) we obtained in our scan, as a function of the supersymmetric contribution to the muon anomalous magnetic moment, $\delta a_\mu$. Filled circles represent models consistent with the current 95\% C.L. range for beyond the standard model contributions to $(g_\mu-2)$, while empty circles denote all other models. As we anticipated, large values of $\delta a_\mu$ do not always imply large $|g^S_{\rm RR}|$, and, vice-versa. The values of $|g^S_{\rm RR}|$ compatible with the limits on $(g_\mu-2)$ and with all constraints on the supersymmetric setup can be as large as a few times $10^{-4}$, though the size of the effect could also be many orders of magnitude smaller. The models giving the largest effects tend to have large L-R mixing (and hence large trilinear scalar couplings) both in the smuon and in the selectron spectrum, and, naturally, a light supersymmetric particle spectrum. In contrast, assuming alignment between the triscalar and Yukawa matrices leads to unobservably small effects in $\mu$-decay.

Current limits on the parameter $g^S_{RR}$ obtained from direct studies of $\mu$-decay observables lead to an upper bound of $0.067$ according to the recent global analysis of Ref.~\cite{Gagliardi:2005fg}. Thus, improvements in precision by more than two orders of magnitude would be required to probe these non-$(V-A)\otimes (V-A)$ contributions in the large L-R mixing regime. On the other hand, the impact of $g^S_{RR}$ on the extraction of $G_\mu$ from the muon lifetime could become discernible at the level of precision of the muon lifetime measurements  underway at PSI\cite{fast,mulan}. These experiments expect to improve the precision on $\tau_\mu$ such that the experimental error in $G_\mu$ is $10^{-6}$. At such a level, a contribution to $\eta$ from  $g^S_{RR}$ of order $10^{-4}$  would begin to be of interest, as per Eq. (\ref{eq:mudecayrate}). In particular, we note that there exist regions of the MSSM parameter space that generate contributions to $\Delta\tau_\mu/\tau_\mu$ as large as a few $\times 10^{-6}$ via the $\eta$ parameter in Eq.~(\ref{eq:mudecayrate}), corresponding to $\sim$ ppm corrections to $G_\mu$. 

Consideration of this correction could be particularly interesting if future measurements at a facility such as GigaZ lead to comparable improvements in other electroweak parameters, such as $M_Z$ and $\sin^2{\hat\theta}_W(M_Z)$. A comparison of these quantities can provide a test of the SM (or MSSM) at the level of electroweak radiative corrections via the relation\cite{Marciano:1999ih}
\begin{equation}
\label{eq:Gfswmz}
\sin^2{\hat \theta}_W (M_Z) \cos^2{\hat \theta}_W (M_Z) = \frac{\pi\alpha}{\sqrt{2} M_Z^2 G_\mu\left[1-\Delta{\hat r}(M_Z)\right]}
\end{equation}
where $\Delta{\hat r}(M_Z)$ contains electroweak radiative corrections to the $(V-A)\otimes(V-A)$ $\mu$-decay amplitude, the $Z$-boson self energy, and the running of ${\hat\alpha}$. Any discrepancy in this relation could signal the presence of new physics contributions to $\Delta{\hat r}(M_Z)$ beyond those obtained in the SM (or MSSM). Inclusion of ppm corrections to $G_\mu$ arising from the presence of a non-zero $\eta$ in Eq.~(\ref{eq:mudecayrate}) would be important in using Eq.~(\ref{eq:Gfswmz}) to carry out a ppm self-consistency test. Resolution of other theoretical issues in the computation of  $\Delta{\hat r}(M_Z)$ -- such as hadronic contributions to the running of $\hat\alpha$ -- would also be essential in performing such a test.

In the case of $\beta$-decay, we show the analogue of the figure described above for the muon decay, in Fig.~\ref{fig:amu}, (b). We show, as a function of $\delta a_\mu$, the value of $a^S_{RR}+a^S_{RL}$. We find that values of $a^S_{RR}+a^S_{RL}$ as large as $10^{-3}$ are consistent with all phenomenological constraints. Since the amplitudes ${\tilde\delta}_\beta^{(a,b)}$ 
depend on L-R mixing among first, rather than second, generation sleptons and squarks (the factors $Z_L^{1i^\prime} Z_L^{4i^\prime\, \ast}$ and $Z_Q^{1i\, \ast} Z_L^{4i}$, $Q=U$ or $D$, respectively) the parameters 
$a^S_{RR}$, $a^S_{RL}$, and $a^T_{RL}$ are not as constrained by precision measurements as is $g^S_{RR}$. Thus, it is possible for the $\beta$-decay parameters to reach their naive, maximal scale $\alpha/4\pi$ in the limit of maximal L-R mixing.

The correlations between $g^S_{RR}$ and $a^S_{RR}$, and between $a^S_{RR}$ and $a^S_{RL}$, are shown in the panels (a) and (b), respectively, of Fig.~\ref{fig:corr}. In the figure, again, filled circles represent models consistent with the current 95\% C.L. range for beyond the standard model contributions to $(g_\mu-2)$, while empty circles denote all other models. We notice that in general there exists no strong correlation among the various quantities, ane we find some correlation only for very large values of the quantities under investigation, in the upper right portions of the plots. Also, no hierarchy between $a^S_{RR}$ and $a^S_{RL}$ exists.

A correlation between the various quantities of interest does arise, however,  when some priors are in place on the supersymmetric particle spectrum. As alluded above, grand unification motivates a matter (sfermion) soft breaking mass sector with some correlations between slepton and squark masses. As an illustrative \lq\lq toy model", we considered a variant of minimal supergravity (mSUGRA), where sfermion soft breaking mass universality is imposed at the GUT scale, to generate a pattern of {\em correlated} slepton and squark soft supersymmetry breaking masses. To generate a large effect in the quantities we study here, we artificially imposed a maximal L-R mixing (we therefore {\em do not} consider our results the outcome of the mSUGRA model, of which we only retain the slepton and squark soft supersymmetry breaking mass pattern), and studied, in Fig.~\ref{fig:enh} (a), the correlation between $a^S_{\rm RR}+a^S_{\rm RL}$ and $g^S_{\rm RR}$.

In particular, we display on the vertical axis the values of $a^S_{RR}+a^S_{RL}$, and on the horizontal axis $g^S_{RR}$, for models where we imposed a maximal L-R mixing. In this case, we obtain that $0.7\lesssim (a^S_{\rm RR}+a^S_{\rm RL})/g^S_{\rm RR}\lesssim 1.5$. In contrast to the model-independent parameter space scans, a nearly linear correlation between these parameters arises due to the mSUGRA-dependent relations between sfermion masses and  requirement of maximal L-R mixing. Moreover, the magnitude of the $\beta$-decay couplings is generally less than $\lesssim 10^{-4}$ due to the $(g_\mu-2)$ constraints on smuon masses and the mSUGRA sfermion mass relations. It is interesting to note that  within this model scenario, the observation of a non-zero $\beta$-decay correlation at the $\sim 10^{-4}$ level would imply a non-zero $g^S_{RR}$ of similar magnitude, along with the corresponding correction to the theoretical $\mu$-decay rate. 

In the more general, model-independent situation, it is important to emphasize that  large L-R mixing in the first generation slepton and squark sectors can lead to $a^S_{RR}$, $a^S_{RL}$, and $a^T_{RL}$ as large  as ${\cal O}(10^{-3})$. Coefficients of this magnitude could, in principle, be probed with a new generation of precision $\beta$-decay correlation studies. At present, the most precise tests of these quantities arises from superallowed Fermi nuclear $\beta$-decay, from which one obtains constraints on the Fierz interference coefficient $b_F = 0.0026 (26)$ \cite{Hardy:2004dm,Hardy:2004id}. For this transition one has
\begin{equation}
b_F = \pm \frac{2\, g_S}{g_V}\, \frac{a^S_{RL}+a^S_{RR}}{a^V_{LL}}
\end{equation}
independent of the details of the nuclear matrix elements\footnote{Here, we have assumed all quantities are relatively real.}. In Fig.~\ref{fig:amu}, (b), we also show the quantity $b_F$ assuming $g_S/g_V=1$ in the right-hand vertical axis.  The present experimental sensitivity lies just above the upper end of the range of possible values of $b_F$.

It is also interesting to consider the recent global analysis of Ref.~\cite{Severijns:2006dr}, where several different fits to $\beta$-decay data were performed. The fit most relevant to the presence analysis corresponds to \lq\lq case 2" in that work, leading to bounds on the following quantities:
\begin{eqnarray}
R_S & \equiv & \frac{ g_S}{g_V}\, \frac{a^S_{RL}+a^S_{RR}}{a^V_{LL}}\\
\nonumber
R_T & \equiv & \frac{ 2\, g_T}{g_V}\, \frac{a^T_{RL}}{a^V_{LL}}
\end{eqnarray}
In particular, including latest results for the neutron lifetime\cite{Serebrov:2004zf} that differs from the previous world average by six standard deviations leads to a non-zero $R_T$: $R_T = 0.0086(31)$ and $R_S=0.00045(127)$ with $\chi^2/{\rm d.o.f.} = 1.75$. In contrast, excluding the new $\tau_n$ result implies both tensor and scalar couplings consistent with zero. We note that SUSY box graphs could not account for tensor couplings of order one percent since the natural scale of the relevant correction -- ${\tilde\delta}_\beta^{(b)}$ -- is $\alpha/2\pi\sim 0.1\%$ in the case of maximal L-R mixing and SUSY masses of order $M_Z$ [see Eq.~(\ref{eq:deltabetab})]. Moreover, there exist no logarithmic or large $\tan\beta$ enhancements that could increase the magnitude of this amplitude over this scale.

Future improvements in experimental sensitivity by up to an order of magnitude could allow one to probe the MSSM-induced non-$(V-A)\otimes(V-A)$ contributions to the $\beta$-decay correlation coefficients in the regime of large L-R mixing. For example, future experiments using cold and ultracold neutrons could allow a determination of the energy-dependent component of the neutrino asymmetry parameter $B$ in polarized neutron decay at the level of a few $\times 10^{-4}$. As indicated by the scatter plot in Fig.~\ref{fig:enh} (b) -- where we show the range of values for the energy-dependent part of the neutrino asymmetry --   experiments with this level of sensitivity could probe well into the region of parameter space associated with large L-R mixing\cite{brad}. Similarly, prospects for significant improvements in the sensitivity to the Fierz interference term using nuclear decays at a new radioactive ion beam facility are under active consideration \cite{savard}.

As with  other low-energy, semileptonic observables, the theoretical interpretation of the $\beta$-decay correlation coefficients requires input from hadron structure theory. For example, the form factors that multiply the scalar and tensor couplings have not been determined experimentally, and there exists some latitude in theoretical expectations for these quantities. The current estimates  are~\cite{Herczeg:2001vk}
\be
\label{eq:ffs}
0.25 \; \lesssim\;  g_S \; \lesssim \; 1 \qquad \qquad \qquad 0.6 \; \lesssim \; g_T \; \lesssim \; 2.3 ;.
\ee
These ranges derive from estimates for neutral current form factors assuming the quark model and spherically-symmetric wavefunctions~\cite{Adler:1975he}. In obtaining the dependence of $b_F$ and $B$ on MSSM parameters as in Figs.~\ref{fig:amu},\ref{fig:enh}, we have assumed $g_S/g_V=1=g_T/g_A$, so the final sensitivities of correlation studies to MSSM-induced non-$(V-A)\otimes(V-A)$ interactions will depend on firm predictions for these ratios.
Similarly, the effects of second class currents generated by the small violation of strong isospin symmetry in the SM may generate $\beta$ energy-dependent contributions to the correlation coefficients that mimic the effects of the MSSM-induced scalar and tensor interactions discussed here. An analysis of these effects on the correlation coefficients $a$ and $A$ has been recently performed in Ref.~\cite{Gardner:2000nk}. To our knowledge, no such study has been carried out for the correlation coefficients of interest here. Carrying out such an analysis, as well as sharpening the theoretical expectations of Eq.~(\ref{eq:ffs}), would clearly be important for the theoretical interpretation of future correlation studies.

\section{Discussion and Conclusions}
\label{sec:conclude}

If supersymmetric particles are discovered at the LHC, it will be then important to draw on measurements of a wide array of observables in order to determine the parameters that describe the superpartners interactions. As we have discussed above, precision studies of weak decay correlations may provide one avenue for doing so. In particular, such studies could probe a unique feature of SUSY not easily accessed elsewhere, namely,  triscalar interactions involving first and second generation scalar fermions. The  presence of triscalar interactions is implied by both purely supersymmetric Yukawa and bilinear components of the superpotential and by soft, SUSY-breaking  triscalar interactions in the Lagrangian. The absolute mass scale, flavor and chiral structure of the triscalar interactions are particularly vexing, since -- in the MSSM -- one has no {\em a priori} reason to take the supesymmetric $\mu$ parameter to be of order the electrowek scale and since the SUSY-breaking interactions introduce both a large number of {\em a priori} unknown parameters and -- experimentally -- a limited number of handles with which to probe them. 

In light of this situation, it has been the common practice to assume that there exists a mechanism forcing $|\mu|\sim v$ and to rely on models that relate various parameters and reduce the number of inputs that must be determined from data.  Conventionally, one makes the \lq\lq alignment" assumption, wherein the soft-triscalar couplings for a given species of fermion are proportional to the corresponding Yukawa matrices. Under this assumption -- along with that of $|\mu|\sim v$ -- one would expect the effects of soft triscalar interactions to be suppressed for the first and second generations.  As we have argued above, the study of weak decay correlations offer a means for experimentally testing the possibility that either $| \mu| >> v$ or $a_f\sim v$. The latter would have particularly interesting implications for Higgs studies at the LHC or ILC, as all but the lightest CP-even, SM-like Higgs would be too heavy to be observed.  

The effects of  triscalar couplings in weak decay correlations arise from one-loop graphs that generate scalar and tensor interactions. These interactions are forbidden in the SM CC interaction in the limit of massless fermions since it involves only LH fermions and since the scalar and tensor operators couple fields of opposite chirality. In the MSSM, such terms can arise via L-R mixing of virtual scalar fermions in one-loop box graphs, and this L-R mixing can be significant when either $\mu$ is large or the corresponding soft, triscalar couplings are unsuppressed. In the case of $\mu$-decay, additional contributions to scalar and tensor four-fermion operators can also be generated by flavor-mixing among same-chirality scalar leptons, but this flavor-mixing is highly constrained by LFV studies such as $\mu\to e\gamma$. Thus, for both $\mu$- and $\beta$-decay, observable, SUSY-induced scalar and tensor couplings can only be generated by flavor diagonal L-R mixing. 

Probing these interactions would require improvements in precision of one- and two-orders of magnitude, respectively,  for $\beta$-decay and $\mu$-decay correlation coefficients. Order of magnitude improvements for $\beta$-decay appear realistic, while the necessary advances for $\mu$-decay appear to be more daunting. On the other hand, if SUSY is discovered at the LHC, then considerations of SUSY-induced, four-fermion scalar interactions involving RH charged leptons may become necessary when extracting the Fermi constant from the muon lifetime. Doing so could become particularly important when ppm tests of electroweak symmetry become feasible.


\vspace*{1cm}
\noindent{{\bf Acknowledgments} } \\
\noindent The authors gratefully acknowledge useful conversations with B.~Filippone, J.~Hardy, R.~Holt,  Z.~T.~Lu, P.~Paradisi and G.~Savard. MR-M thanks A.~Kurylov with whom part of the computation was carried out in collaboration.
This work was supported in part U.S. Department of Energy contracts FG02-05ER41361 and DE-FG03-92-ER40701,  N.S.F. award PHY-0555674, and NASA contract NNG05GF69G.


\end{document}